\documentclass[iop, numberedappendix]{emulateapj}

\usepackage{apjfonts}

\usepackage{color}
\definecolor{citeRGB}{rgb}{0,0.1,0.7}
\usepackage[hyperfootnotes=true,naturalnames=true,letterpaper,pdfstartview=FitH,pdfpagemode=UseNone,colorlinks=true,citecolor=citeRGB]{hyperref}

\usepackage{etoolbox}

\makeatletter

\patchcmd{\NAT@citex}
  {\@citea\NAT@hyper@{\NAT@nmfmt{\NAT@nm}\NAT@date}}
  {\@citea\NAT@nmfmt{\NAT@nm}\NAT@hyper@{\NAT@date}}
  {}
  {}

\patchcmd{\NAT@citex}
  {\@citea\NAT@hyper@{%
     \NAT@nmfmt{\NAT@nm}%
     \hyper@natlinkbreak{\NAT@aysep\NAT@spacechar}{\@citeb\@extra@b@citeb}%
     \NAT@date}}
  {\@citea\NAT@nmfmt{\NAT@nm}%
   \NAT@aysep\NAT@spacechar%
   \NAT@hyper@{\NAT@date}}
  {}
  {}

\patchcmd{\NAT@citex}
  {\@citea\NAT@hyper@{%
     \NAT@nmfmt{\NAT@nm}%
     \hyper@natlinkbreak{\NAT@spacechar\NAT@@open\if*#1*\else#1\NAT@spacechar\fi}%
       {\@citeb\@extra@b@citeb}%
     \NAT@date}}
  {\@citea\NAT@nmfmt{\NAT@nm}%
   \NAT@spacechar\NAT@@open\if*#1*\else#1\NAT@spacechar\fi%
   \NAT@hyper@{\NAT@date}}
  {}
  {}

\makeatother

\gdef\HST{\textit{HST}}
\gdef\G141{\textit{G141}}
\gdef\F140W{\textit{F140W}}
\gdef\fluxcgs{\mathrm{erg\ s^{-1}\ cm^{-2}}}
\gdef\micront{$\mu$m}
\gdef\micronm{\mu\mathrm{m}}

\gdef\flux_radius{\textsc{flux\_radius}}

\gdef\lensID{SL2SJ02176-0513}

\gdef\logOH{12 + \log \left(\mathrm{O/H}\right)}

\gdef\24mum{$24\,\mu\mathrm{m}$}

\gdef\4ang{4000\,\AA}

\gdef\galfit{\texttt{galfit}}
\gdef\emcee{\texttt{emcee}}

\shortauthors{Brammer et al.}
\shorttitle{\HST\ observations of a $z=1.847$ lens}
\slugcomment{ApJL accepted (\today)}

\begin{document}

\title{3D-HST Grism Spectroscopy of a Gravitationally Lensed, Low-metallicity Starburst Galaxy at $z=1.847$\footnotemark[*]}

\author{Gabriel~B.~Brammer\altaffilmark{1},
Rub\'en S\'anchez-Janssen\altaffilmark{1}, 
Ivo Labb\'e\altaffilmark{2},
Elisabete da Cunha\altaffilmark{3},
Dawn K.\ Erb\altaffilmark{4},
Marijn Franx\altaffilmark{2},
Mattia Fumagalli\altaffilmark{2},
Britt Lundgren\altaffilmark{5},
Danilo Marchesini\altaffilmark{6},
Ivelina Momcheva\altaffilmark{5},
Erica Nelson\altaffilmark{5},
Shannon Patel\altaffilmark{2},
Ryan Quadri\altaffilmark{7},
Hans-Walter Rix\altaffilmark{3},
Rosalind E.\ Skelton\altaffilmark{5},
Kasper B.\ Schmidt\altaffilmark{3},
Arjen van der Wel\altaffilmark{3},
Pieter G.\ van Dokkum\altaffilmark{5},
David A.\ Wake\altaffilmark{5},
Katherine E.\ Whitaker\altaffilmark{5},
}

\email{gbrammer@eso.org}

\altaffiltext{1}
{European Southern Observatory, Alonso de C\'ordova 3107, Casilla 19001, Vitacura, Santiago, Chile}
\altaffiltext{2}
{Leiden Observatory, Leiden University, Leiden, The Netherlands}
\altaffiltext{3}
{Max Planck Institute for Astronomy (MPIA), K\"onigstuhl 17,
69117, Heidelberg, Germany}
\altaffiltext{4}
{Department of Physics, University of Wisconsin-Milwaukee, P.O. Box 413,
Milwaukee, WI 53201, USA}
\altaffiltext{5}
{Department of Astronomy, Yale University, New Haven, CT 06520, USA}
\altaffiltext{6}
{Physics and Astronomy Department, Tufts University, Robinson Hall,
Room 257, Medford, MA, 02155, USA}
\altaffiltext{7}
{Carnegie Observatories, 813 Santa Barbara Street, Pasadena, CA 91101, USA}

\footnotetext[*]{Based on observations made with the NASA/ESA \textit{Hubble Space Telescope}, program \#12328, obtained at the Space Telescope Science Institute, which is operated by the Association of Universities for Research in Astronomy, Inc., under NASA contract NAS 5-26555.}

\begin{abstract}

We present \textit{Hubble Space Telescope} (\HST) imaging and spectroscopy of the gravitational lens SL2SJ02176-0513, a cusp arc at $z=1.847$.  The UV continuum of the lensed galaxy is very blue, which is seemingly at odds with its redder optical colors.  The 3D-HST WFC3/G141 near-infrared spectrum of the lens reveals the source of this discrepancy to be extremely strong [\ion{O}{3}]$\lambda$5007 and H$\beta$ emission lines with rest-frame equivalent widths of $2000\pm100$ and $520\pm40$~\AA, respectively.  The source has a stellar mass $\sim10^8~M_\odot$, $sSFR\sim100/\mathrm{Gyr}$, and detection of [\ion{O}{3}]$\lambda$4363 yields a metallicity of $\logOH=7.5\pm0.2$.  We identify local blue compact dwarf analogs to \lensID, which are among the most metal-poor galaxies in the SDSS.  The local analogs resemble the lensed galaxy in many ways, including UV/optical SED, spatial morphology and emission line equivalent widths and ratios.  Common to \lensID\ and its local counterparts is an upturn at mid-IR wavelengths likely arising from hot dust heated by starbursts.  The emission lines of \lensID\ are spatially resolved owing to the combination of the lens and the high spatial resolution of \HST.  The lensed galaxy is composed of two clumps with combined size $r_e\sim$300~pc, and we resolve significant differences in UV color and emission line equivalent width between them.  Though it has characteristics occasionally attributed to active galactic nuclei, we conclude that \lensID\ is a low-metallicity star-bursting dwarf galaxy.  Such galaxies will be found in significant numbers in the full 3D-HST grism survey. 

\end{abstract}

\keywords{galaxies: formation --- galaxies: starburst --- galaxies: dwarf --- galaxies: high-redshift}

\section{Introduction}
\label{s:introduction}

Tracing galaxy populations over cosmic time is key to understanding their evolutionary processes.  Observations of a particular galaxy type over a range of redshifts yield complementary information, from spatially-resolved, high signal-to-noise studies of individual local examples to more distant statistical samples \citep[e.g.,][]{overzier:09}.  While local low-mass dwarfs ($\lesssim10^9~M_\odot$) can be studied relatively easily, more distant examples are usually missing in typical flux-, mass-, or star-formation-rate-limited surveys.  The best-studied examples at cosmological distances are usually strongly magnified by gravitational lenses \citep[e.g.,][]{fosbury:03, yuan:09, ewuyts:12}.  

The lowest mass galaxies tend to have the lowest metallicities \citep{tremonti:04}, and the normalization of this mass-metallicity relation evolves such that galaxies at a given stellar mass have lower metallicities at higher redshifts to at least $z\sim3$ \citep{erb:06, mannucci:09}.  The physical conditions of low-metallicity star-forming galaxies can be very different than their higher-mass counterparts, characterized by a hard radiation field and strong stellar winds \citep[e.g.,][]{erb:10}.  Consequently they may have strong emission lines, which can be used to select them \citep{atek:11, vanderwel:11} and also must be accounted for when modeling their broad-band photometry \citep{atek:11, finkelstein:11}.  Active galactic nuclei (AGN) are also sources of hard ionizing radiation that can result in similar spectral features, so accurately separating the influence of AGN and metallicity/star formation is crucial for understanding the dominant processes that shape galaxies at $z>1$ \citep[e.g.,][]{trump:11}.

In this Letter, we present \HST\ observations of a gravitational lens system discovered by the Strong Lensing Legacy Survey consisting of a bright cusp arc at $z=1.8470$ that is lensed by a massive galaxy at $z=0.6459$ \citep[SL2SJ02176-0513,][]{tu:09}.  The system lies within the CANDELS imaging \citep{grogin:11, koekemoer:11} and 3D-HST spectroscopic \citep{brammer:3dhst} surveys of the UKIDSS/UDS field.  We use the unique combination of the \HST\ datasets and the natural lens to demonstrate that the lensed galaxy is a low-mass, low-metallicity galaxy undergoing an extreme starburst, and that it shares many characteristics of local low-metallicity blue compact dwarf galaxies.  We adopt cosmological parameters $h=0.7$, $\Omega_m=0.3$, and $\Omega_\Lambda=0.7$ throughout.

\section{Observations}
\label{s:observations}

\begin{figure}
\epsscale{0.95}
\plotone{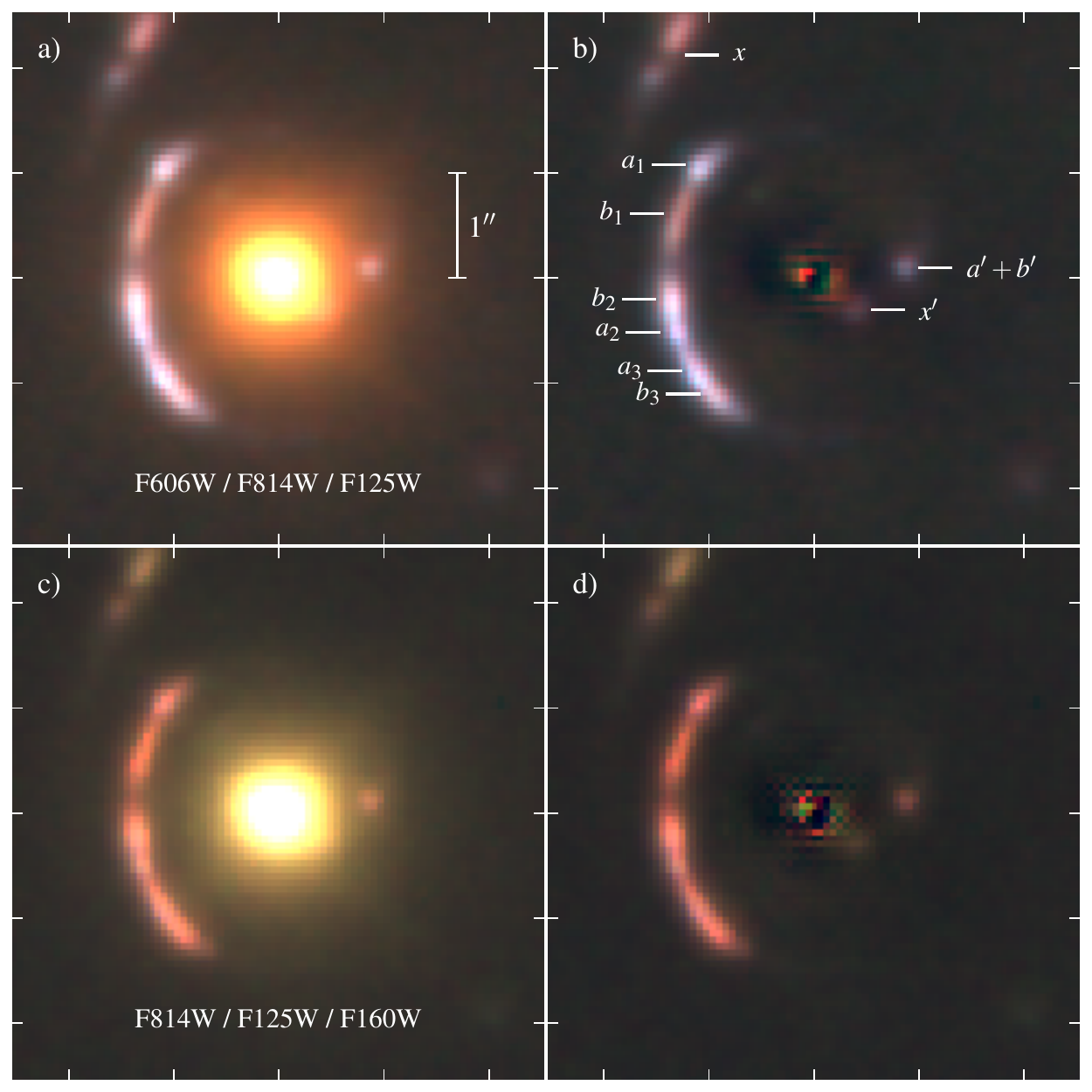}
\caption{Three-color images of the UDS lens + arc system.  Panels $a)$ and $b)$ are made from ACS F606W (blue), F814W (green) and WFC3 F125W (red).  Panels $c)$ and $d)$ are made from ACS F814W (blue), WFC3 F125W (green), and F160W (red).  All images are PSF-matched to F160W.  The band-dependent \galfit\ model of the foreground lens galaxy is subtracted from the right panels.   The object lensed into the cusp arc is composed of two primary clumps, ``$a$'' and ``$b$,'' and the conjugate image positions are shown in panel $b)$.  The conjugate images of an additional lensed object  \citep[][]{tu:09, cooray:11} are labeled ``$x$.'' \label{f:three_color}}  
\end{figure}

We use the ``v1.0'' reduction of the 1--2 orbit \HST\ ACS F606W, F814W and WFC3 F125W and F160W images provided by the CANDELS team \citep{koekemoer:11}.  Three-color combinations of the CANDELS images are shown in Figure~\ref{f:three_color}.  The arc is very blue through the observed optical and F125W bands, with the UV slope $\beta=-1.7\pm0.2$ ($f_\lambda\propto\lambda^\beta$) determined from the two ACS bands ($\lambda_\mathrm{rest}=$2100--2800\AA).  The arc becomes surprisingly red when the F160W band is included; \cite{cooray:11} hypothesized that the red F125W$-$F160W color is caused by strong [\ion{O}{3}] emission that dominates the flux in the redder band, similar to the extreme equivalent-width galaxies discovered by \cite{vanderwel:11}.  In addition to the \HST\ observations, we extract photometry of the arc from CFHT-LS $ugriz$ and UKIDSS/UDS $JHK$ imaging following \cite{cooray:11} and removing the roughly symmetric lens galaxy by subtracting a flipped version of the ground-based images.

\begin{figure}
\epsscale{1.2}
\plotone{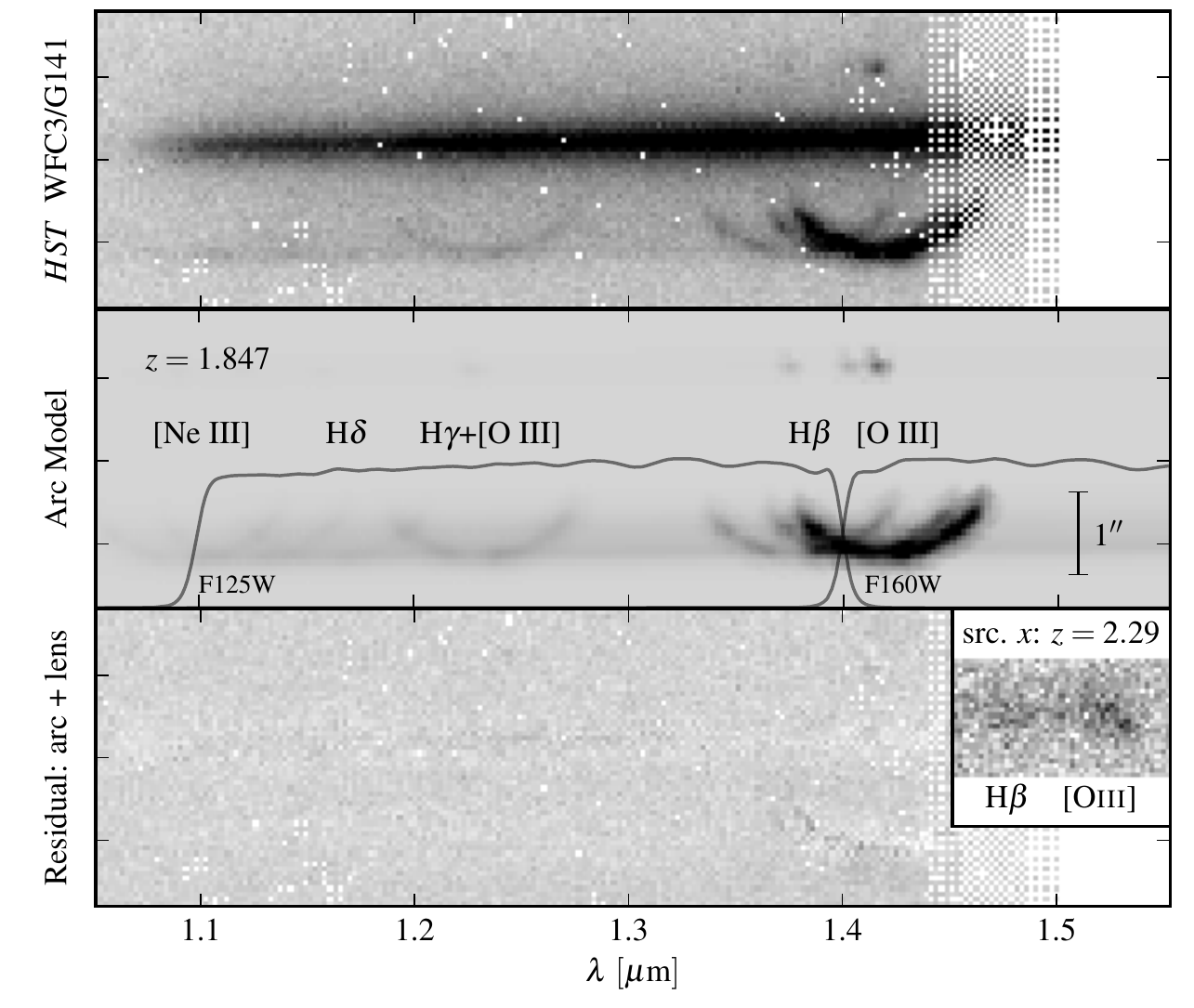}
\caption{Two-dimensional slitless WFC3/G141 spectrum of the UDS lens + arc system from 3D-HST.  \textbf{Top:} The combined spectrum of the $z=0.656$ lens galaxy and the arc. The dispersion axis lies roughly along the arc resulting in a clean separation of the arc and lens but also in overlapping emission lines.  \textbf{Middle:} The modeled line+continuum two-dimensional spectrum.  The G141 spectrum is cut off at the edge of the 3D-HST pointing; the pixelated structure at $\sim$1.45\micront\ is the result of the dither pattern.  \textbf{Bottom:} Residuals of the lens + arc model fit.  The small inset panel shows detections from an adjacent 3D-HST pointing of weak H$\beta$ and [\ion{O}{3}] emission lines for the lensed source ``$x$'' (Figure \ref{f:three_color}) at $z=2.29$ . \label{f:twod_spectrum}}  
\end{figure}

\lensID\ was observed with the WFC3 G141 grism on 2011 December 21 as part of the 3D-HST survey.  The spectrum was reduced as described in detail by \cite{brammer:3dhst}.  Figure~\ref{f:twod_spectrum} shows the remarkable G141 spectrum of the lens system, with strong emission lines of [\ion{O}{3}]$\lambda$4959+5007, H$\beta$, and H$\gamma$ (potentially blended with [\ion{O}{3}]$\lambda$4363) that explain the colors in Figure~\ref{f:three_color} and confirm the prediction of \cite{cooray:11}.  All of the emission lines are extended, showing the same morphology as the UV (ACS) continuum.  The emission lines prove that the arc and counter image lie at the same redshift.

The \lensID\ system is a bright MIPS 24\micront\ source\footnote{http://irsa.ipac.caltech.edu/data/SPITZER/SpUDS} with a total flux of 0.565 mJy, offset from the lens galaxy and centered on the arc.  From a simultaneous fit \citep[see][]{labbe:06} of the 24\micront\ contributions from the arc, the lens, and the additional $z=2.3$ lensed galaxy to the north, we find that the arc contributes 85\% of the 24\micront\ flux, or $0.476\pm0.025$ mJy.

\section{Analysis}
\label{s:analysis}

\subsection{Lens magnification}
\label{s:lens_magnification}

Below we determine physical properties of \lensID\ integrated over the full arc, which is composed of three magnified images of the source (Figure~\ref{f:three_color}).  To estimate an upper limit on the true physical scales of the source, we consider the brightness and size of the faint counter image opposite of the arc. We determine a brightness ratio $\mu_\star=25$ between the integrated arc and the counter image.   The magnification of the counter image, $\mu^\prime$, is $\sim$1.4 given the lens model parameters of \cite{cooray:11}.  The counter image, which includes both clumps resolved in the arc, has $\sqrt{\mu^\prime}\cdot r_e=1.4~\mathrm{pix}=0\farcs04=350~\mathrm{pc},~r_e\sim300~\mathrm{pc}$ in the ACS images.

\subsection{Modeling the grism spectrum}
\label{s:make_model}

In order to extract quantities from the grism spectrum, we generate and fit a model of the two-dimensional spectrum that essentially convolves an arbitrary one-dimensional spectrum with an assumed object morphology, given the grism dispersion configuration files provided by STScI.  To model the contribution of the lens to the flux at the location of the arc, we adopt as the lens morphology an analytical Sersic profile with parameters determined by running \galfit\ \citep{peng:02} on the 3D-HST F140W image.  For the arc, we adopt the observed (lens-subtracted) F140W morphology and a one-dimensional spectrum that consists of a $Z=0.008$ \cite{bc:03} single stellar population model for the continuum and individual emission lines.  

The model is fit to the observed spectrum with parameters optimized by the \emcee\ Markov Chain Monte-Carlo sampler \citep{emcee}, where the free parameters are 1) a spatial shift and a spectral scaling to improve the subtraction of the lens, 2) the redshift of the arc, 3) the age, stellar mass, and reddening of the arc continuum following a \cite{calzetti:00} reddening law, and 4) individual strengths of the arc emission lines [\ion{O}{3}]$\lambda$4959+5007, H$\beta$, H$\gamma$, [\ion{O}{3}]$\lambda$4363, H$\delta$, H$\epsilon$, and \ion{Ne}{3}$\lambda 3869$.  While the spectral resolution of the G141 grism ($R\sim130$) is insufficient to resolve the H$\gamma$ and [\ion{O}{3}]$\lambda$4363 lines, the \textit{sum} of these lines is well-constrained by the G141 spectrum.  

The best-fit model of the 2D arc spectrum is shown in the middle panel of Figure~\ref{f:twod_spectrum}, and the residuals of the full lens+arc model fit are shown in the bottom panel.  The parameters of the model, including the line strengths and observed-frame equivalent widths are summarized in Table~\ref{t:full}.  The uncertainties on all parameters come from the full marginalized posterior distribution function of the MCMC chain.  The most robust fits are for the $\lambda$4959+5007 and H$\beta$ emission line strengths, with rest-frame equivalent widths of $2000\pm100$ and $520\pm40$ \AA, respectively.  We obtain a marginal detection of  [\ion{O}{3}]$\lambda$4363 at $5\pm3\times10^{-17}~\fluxcgs$ assuming an intrinsic Balmer line ratio H$\gamma$/H$\beta=0.468$ and no reddening of the Balmer lines.  The [\ion{O}{3}]$\lambda$4363 flux will be higher for any non-zero Balmer decrement.  

\begin{deluxetable}{lr}
\tablecolumns{4} 
\tablewidth{0pt} 
\tablecaption{Observables and derived quantities}
\tablehead{\colhead{Parameter}          &
         \colhead{Value}}
\startdata
R.A., Dec.  & 02:17:37.237,\,\,$-$05:13:29.78 \\
\multicolumn{2}{c}{\textit{Photometry}}\\
F606W ($V$) [AB] & $22.07\pm0.03$ \\
F814W ($i$) & $21.96\pm0.03$ \\
F125W ($J$) & $21.71\pm0.02$ \\
F140W ($H_\mathrm{wide}$) & $21.10\pm0.04$\\
F160W ($H$) & $20.90\pm0.02$ \\
MIPS 24\micront\ [mJy] &~$0.476\pm0.025$ \\
$\mu_\star$ & 25$\pm$1\tablenotemark{a} \\
$\mu^\prime$ & $\sim$1.4\tablenotemark{a} \\
\multicolumn{2}{c}{\textit{Spectrum + SED fit}}\\
    $z$ & $   1.84691 \pm    0.0004^\mathrm{rand} \pm 0.002^\mathrm{sys}$ \\
 $\log \left(\mathrm{age/yr}\right)$ & $  7.2 \pm    0.2$ \\
  $A_V$ (continuum) & $0.09 \pm 0.15 $ \\
 $\log \left(\mu\cdot M/M_\odot\right)$ & $   9.5 \pm  0.1 $ \\
$f$([\ion{O}{3}]$\lambda\lambda 4959+5007$) & $309 \pm   2\tablenotemark{b}$ \\
                        $f(\mathrm{H}\beta)$ & $74\pm2$\tablenotemark{b}  \\
  $f(\mathrm{H}\gamma~+~$[\ion{O}{3}]$\lambda 4363)$ & $40\pm3$\tablenotemark{b} \\
                        $f(\mathrm{H}\delta)$ & $13\pm2$\tablenotemark{b} \\
           $f$(\ion{Ne}{3}$\lambda 3869$) & $30\pm4$\tablenotemark{b} \\
           $EW_\mathrm{O~III}$          & $5690 \pm 290$ \AA\tablenotemark{c} \\
           $EW_\mathrm{H\beta}$            & $1470 \pm 110$ \AA\tablenotemark{c} \\
\multicolumn{2}{c}{\textit{Derived parameters}}\\
$\beta$ (2000--2800\,\AA)     & $-1.7 \pm 0.2$ \\
$\mu\cdot SFR_\mathrm{H\beta}\ (M_\odot~\mathrm{yr}^{-1})$ & $390\pm9$  \\
$\logOH$ & $7.5\pm0.2$ \\
$\sqrt{\mu^\prime}\cdot r_e$ & 350 pc  \\
$\Sigma_\mathrm{SFR}~(r<r_e,~M_\odot~\mathrm{yr}^{-1}~\mathrm{kpc}^{-2})$ & 20 \\
\enddata
\tablenotetext{a}{$\mu_\star$ is the relative magnification between the integrated arc and the counter image.  $\mu^\prime$ is the brightness magnification of the counter image.  The total lens magnification is $\mu=\mu_\star\cdot\mu^\prime$.}
\tablenotetext{b}{Line fluxes are in $10^{-17}\,\fluxcgs$, uncorrected for lens magnification.}
\tablenotetext{c}{Observed frame}
\label{t:full}
\end{deluxetable}

\subsection{Fundamental quantities: star-formation rate, stellar mass, and metallicity}
\label{s:fundamental_quantities_star_formation_rate_stellar_mass_metallicity}

The luminosity of the H$\beta$ line corresponds to a de-magnified star-formation rate of $11~M_\odot~\mathrm{yr}^{-1}$ (H$\alpha$/H$\beta$=2.86 \citealp{dopita+sutherland}; \citealp{kennicutt:98}).  The H$\gamma$/H$\beta$ ratio is consistent with $E(B-V)=0$.  The de-magnified stellar mass determined from the combined fit to the UV+optical SED and G141 spectrum is $9\times10^{7}~M_\odot$, similar to the average stellar mass of the equivalent-width-selected galaxies from \cite{vanderwel:11}.  The specific star-formation rate of \lensID\ is extremely high (as expected given the extreme line equivalent widths), with a mass doubling time of only 9 Myr ($sSFR\sim100~\mathrm{Gyr}^{-1}$). 

The (implied) detection of the [\ion{O}{3}]$\lambda$4363 line suggests that \lensID\ has a low metallicity as that line can only form in hot \ion{H}{2} regions with little metal-line cooling.  To estimate the metallicity, we draw values of the line fluxes and $A_V$ from the full marginalized PDF of the MCMC fit.  Assuming that the continuum $A_V$ also attenuates the Balmer lines, a given $A_V$ determines the relative fluxes of H$\gamma$ and [\ion{O}{3}]$\lambda$4363 given the observed H$\beta$ flux and a Balmer decrement.  Combining the observed lines with a random uniform distribution [\ion{O}{3}]$\lambda$5007/[\ion{O}{2}]$\lambda$3727=[2.5,9] \citep{erb:10, atek:11}, we use the prescription of \cite{izotov:06} and determine $T$(\ion{O}{3})$=17000\pm3000~\mathrm{K}$ and $\logOH=7.5$, or $\sim$6\% of the solar value \citep[$\logOH=8.69$, ][]{asplund:09}.  The posterior probability on $\logOH$ has (16,~84) percentiles
  of (7.3,~7.7), but with an extended tail towards higher metallicities ($P=5\%$ for $\logOH > 8$) given the low significance of the [\ion{O}{3}]$\lambda$4363 detection.

\begin{figure*}
\epsscale{1.15}
\plotone{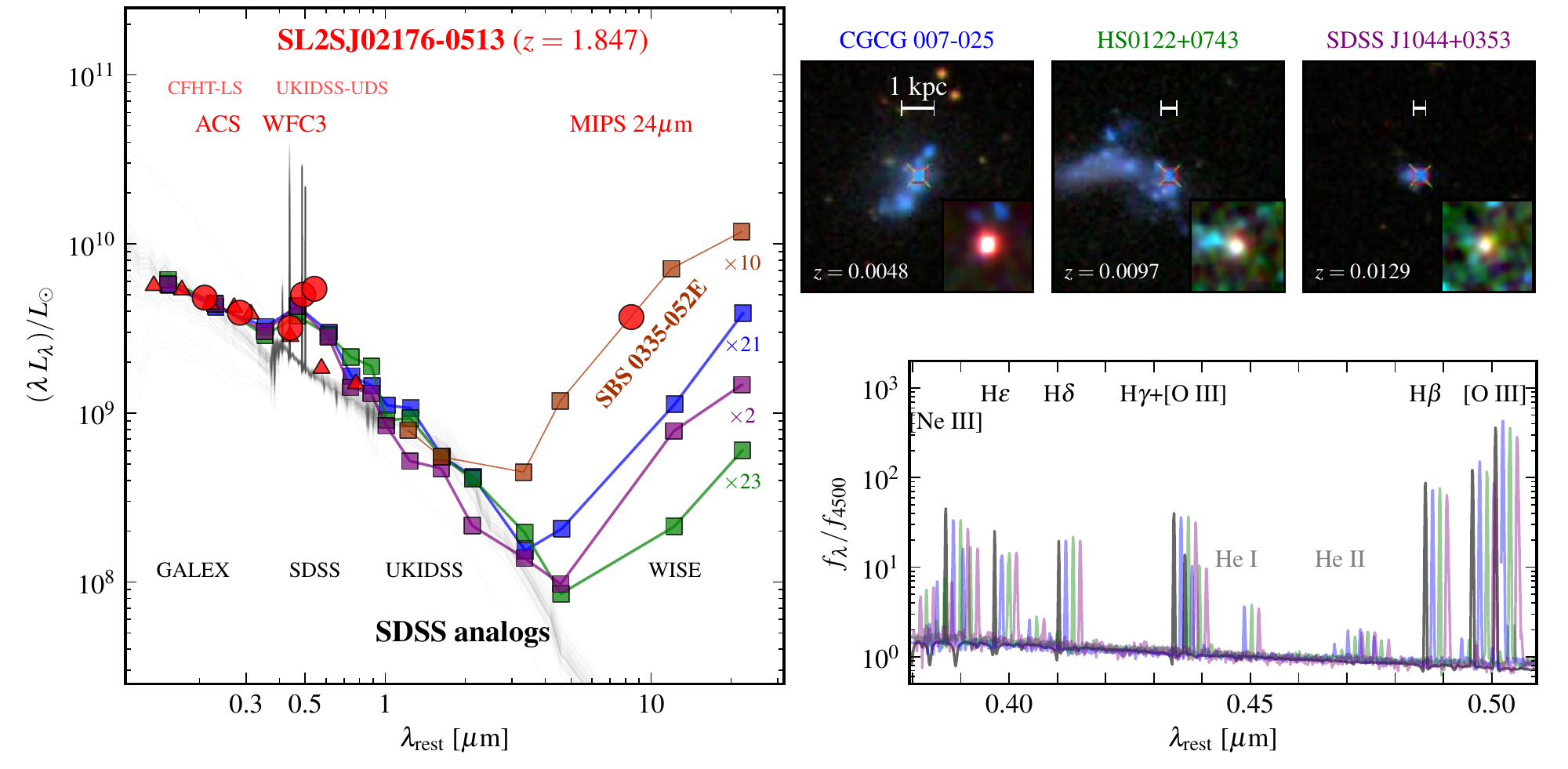}
\caption{Comparison of the \lensID\ SED to local low-metallicity dwarf galaxies from the SDSS.  The left panel shows the integrated photometry of the arc (red symbols; circles space-based, triangles ground-based) and model spectra (gray) drawn from the fit to the photometry and HST grism spectrum. The scaled UV---IR photometry (with factors as indicated) of the local analogs is shown by the squares.  The IR SED of the local metal-poor galaxy SBS~0335-052E is shown in brown, taken from the public 2MASS and WISE catalogs.  The optical morphologies of the SDSS analogs are shown at right, and three-color WISE images are shown in smaller insets.  The thumbnails are all $1\farcm2$ on a side; physical scales of 1 kpc are shown.  The location of the SDSS fiber is shown in the cross; the photometry was extracted from apertures drawn around the entire galaxy.  The lower right panel shows the modeled \lensID\ spectrum (gray line) along with the scaled SDSS spectra of the analogs shifted by 15, 30, and 45~\AA.  The analog spectra and SEDs are colored by object according to the thumbnail labels.  The lines included in the G141 model are labeled in black, while He lines observed in the analogs but too faint to be constrained by the grism spectrum are labeled in gray.\label{f:compare_analogs}}  
\end{figure*}

\section{Local analogs: low-metallicity blue compact dwarf galaxies}
\label{s:local_analogs_low_metallicity_blue_compact_dwarf_galaxies}

While the unique observational data of \lensID\ paint a self-consistent picture of an extremely low-metallicity, low-mass starburst galaxy, many of the simplifying assumptions underlying its derived properties can result in significant systematic uncertainties.  To place the properties of \lensID\ in context with nearby galaxies, we select a comparison sample from the Sloan Digital Sky Survey \citep[SDSS DR7, ][]{sdss:dr7} with $EW_\mathrm{O~III} > 1500$~\AA~ and $\log \left(f_\mathrm{O~III}/f_{\mathrm{H}\beta}\right) < 0.75$\footnote{[\ion{O}{3}]~$\equiv$~[\ion{O}{3}]$\lambda$4959+[\ion{O}{3}]$\lambda$5007} as measured by \cite{brinchmann:04}.  

This simple selection criterion reduces the full SDSS spectroscopic sample of 930~000 galaxies to just 14.  Of these, eight objects have 
$\logOH > 8$ \citep[within the fiber aperture, ][]{tremonti:04} and are in fact compact, especially blue, individual \ion{H}{2} regions in spiral/irregular galaxies (M101 and M106 among them).  The remaining six objects are blue compact dwarf galaxies with $\logOH < 8$.  Four are found in the extremely metal-poor galaxy compilation by \cite{morales-luis:11} with $\logOH\sim7.6$ and the final two have $\logOH\sim7.8$ \citep{brinchmann:08, engelbracht:08}.

\subsection{Spectral energy distributions}
\label{s:spectral_energy_distributions}

Three of these local analog galaxies are compared to \lensID\ in Figure~\ref{f:compare_analogs}.  The left panel demonstrates the similarity of the SEDs of \lensID\ and its local analogs at all observed wavelengths: all have similar blue UV continua, and the optical/NIR photometry of the analogs agrees well with the extrapolation of the lens stellar continuum fit.  \lensID\ is 2--20 times more luminous than the local analogs shown, qualitatively consistent with the order of magnitude increase of stellar mass at constant metallicity to $z\sim2$ \citep[e.g.,][]{erb:06}.  Both samples show a marked upturn at mid-IR wavelengths sampled by WISE \citep{wright:wise} for the local galaxies and by \textit{Spitzer}-MIPS for \lensID.  

Hot dust in low-metallicity blue compact dwarf galaxies has been known and studied for some time \citep[e.g.,][]{thuan:99, hirashita:04, engelbracht:08} and has even recently been used to select new examples with red WISE colors \citep{griffith:11, izotov:11}.  The dust can be produced rapidly by Type II supernovae even in primordial bursts \citep[e.g.,][]{todini:01} and it can be heated by the substantial UV radiation of the intense starbursts \citep{izotov:11}.  

There is significant variation in the IR properties of the local galaxies whose colors and spectral properties are similar in the UV/optical.  While they are rare, examples exist of local galaxies with such rapidly-rising mid-IR dust SEDs as \lensID.  The SED of one such ``mid-IR peaker'' \citep{engelbracht:08} is shown in Figure~\ref{f:compare_analogs} and agrees well with the \lensID\ SED.  This object, SBS 0335-052E, is one of the most metal poor galaxies known in the local universe \citep[$\logOH=7.23\pm0.01$,][]{izotov:97, izotov:09}.  


\begin{figure*}
\epsscale{1.1}
\plotone{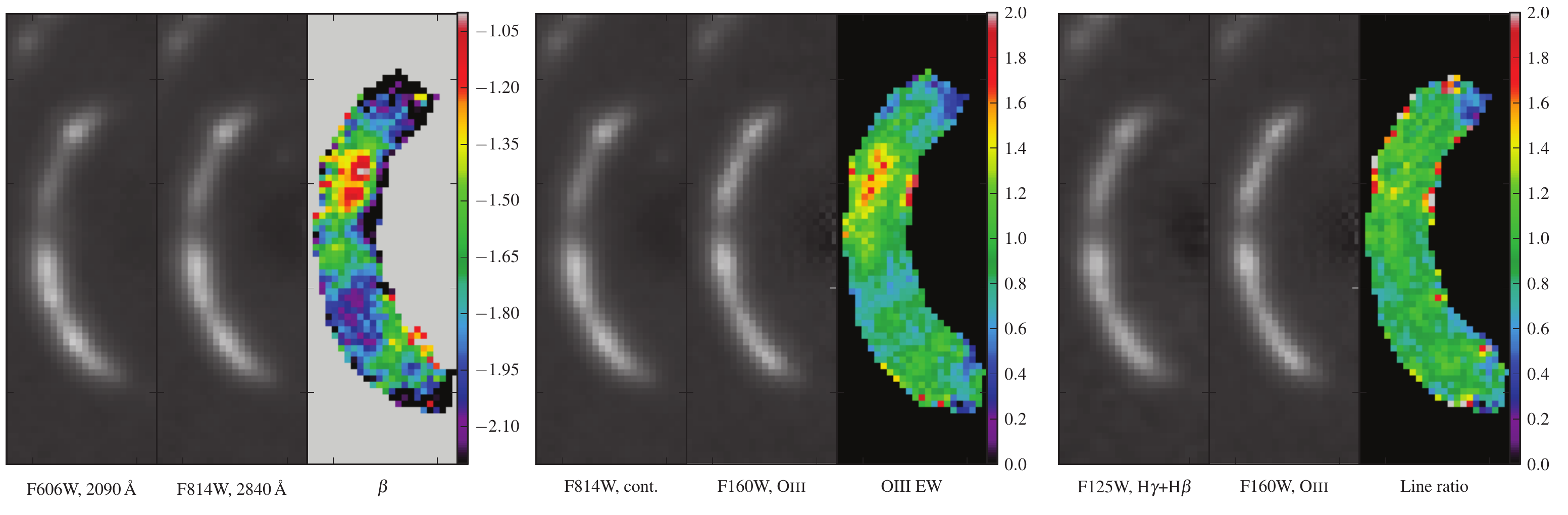}
\caption{Ratios of aligned ACS and WFC3 images.  The \galfit model of the foreground lens has been subtracted from all images and the residual images in the different bands are PSF-matched to the F160W PSF.  \textbf{Left:} Ratio of the ACS F606W and F814W images, combined to provide a map of the UV slope, $\beta$, where $F_\lambda\propto\lambda^\beta$.  \textbf{Center:} Ratio of the WFC3 F160W image, which is dominated by the [\ion{O}{3}] emission lines, to F814W, which is taken to represent the continuum.  The result is a map of the emission line equivalent width across the arc.  \textbf{Right:} The ratio of WFC3 F160W and F125W roughly indicates spatial variations in the [\ion{O}{3}] to Balmer emission line ratio across the arc.  The line ratio is much more uniform than the line equivalent widths.  \label{f:image_ratios}}  
\end{figure*}

\subsection{Emission lines}
\label{s:emission_lines}

The lower-right panel of Figure~\ref{f:compare_analogs} compares the model spectrum of \lensID\ and the SDSS spectra of the local analogs.  All of the spectra shown are qualitatively similar and the relative line strengths agree to within $\lesssim$20--30\%.  [\ion{O}{3}]$\lambda$4363 is detected in all of the of the low-metallicity analogs.  The high ionization \ion{He}{2}$\lambda${4686} line is detected in the SDSS spectra but is too weak to be seen in the arc grism spectrum.  However, \ion{He}{2}$\lambda${1640} is detected in a Keck LRIS spectrum of \lensID\ that was first used to measure the redshift of the arc \citep{tu:09}.  Additional high-ionization UV lines of \ion{C}{4}, \ion{O}{3}], \ion{N}{3}], and \ion{C}{3}] are detected in the arc, prompting \citep{tu:09} to speculate that they are excited by an AGN.  These rest-UV lines have been detected in other distant star-forming galaxies: an H$\alpha$ emitter at $z=2.5$ \citep{malkan:96}, the Lynx arc at $z=3.4$ \citep{fosbury:03}, and a unique Lyman-break galaxy at $z=2.3$ \citep{erb:10}.  \cite{erb:10} argue that the low ratio (\ion{C}{4}/\ion{C}{3}])$\sim$0.3 in that galaxy ($\sim$0.5 in the arc) disfavors a significant contribution of an AGN to the ionizing flux (\ion{C}{4}/\ion{C}{3}]$\sim$1.2, \citealp{hainline:11}).  
 
\subsection{Morphology}
\label{s:morphology}

The optical morphologies of three of the SDSS analog galaxies are shown in the thumbnails of Figure~\ref{f:compare_analogs}.  All of the analogs, like \lensID, are dominated by multiple bright blue clumps with sizes of a few hundred parsecs.  SBS 0335-052E is also composed of multiple ``super star clusters'' with ages 3--15 Myr contained within a scale of 500 pc \citep{thuan:97, reines:08}.  Since all of the individual clumps share the properties of young bursts, the star-formation must be somewhat synchronized across the entire galaxy.  

In Figure~\ref{f:image_ratios} we take different combinations of the \HST ACS and WFC3 images of \lensID\ to compare the properties of the two main clumps resolved in the arc.  The ratio F606W/F814W provides a rough map of the UV-slope across the arc, and we find that the source ``$b$'' is significantly redder than source ``$a$'' (Figure~\ref{f:three_color}; the color image itself shows this difference).  The redder source has an [\ion{O}{3}] equivalent width $\sim$50\% higher than the bluer source in the ratio of the F160W (line) and F814W (continuum) images, however, the F160W ([\ion{O}{3}]) to F125W (Balmer line)  ratio is roughly constant across the arc.  The differences between the sources is most apparent for the $a_1$ and $b_1$ components that are best separated by the lens, though the other pairs of conjugate images show roughly the same trends.  The individual super star clusters in SBS 0335-052E show differences in broad-band colors and line ratios comparable to those in Figure~\ref{f:image_ratios} \citep{thuan:97, izotov:09}.  

\section{Discussion and Summary}
\label{s:discussion_and_summary}

We take advantage of the unique combination of a natural gravitational lens and high spatial resolution \HST\ imaging and near-IR spectroscopy to extract the detailed properties of \lensID.  The near-IR spectrum is dominated by extremely strong emission lines of H$\gamma$, H$\beta$ and [\ion{O}{3}] at $z=1.847$.  From the UV/optical spectrum and photometry, we determine that the source of \lensID\ is a young starbursting ($sSFR\sim100/\mathrm{Gyr}$) dwarf galaxy ($M\sim1.3\times10^{8}~M_\odot$) with an extremely low gas-phase metallicity ($\logOH\sim7.5$).  Even with so few metals, \lensID\ shows detectable hot dust emission observed at $\lambda_\mathrm{obs}=24~\micronm$.  

We find the unique properties of \lensID\ to be remarkably similar to those of nearby low-metallicity blue compact dwarf galaxies selected to have similar [\ion{O}{3}] equivalent widths and ([\ion{O}{3}]/H$\beta$) line ratios.  Such galaxies are frequently considered to be local analogs to galaxies expected to be more common at earlier cosmic times and in \lensID\ we have discovered a compelling connection at high redshift (which itself is likely a bright, magnified example of the galaxies discovered by \citealp{vanderwel:11}).  

\lensID\ has characteristics that are frequently attributed to AGN at similar redshifts:  extreme equivalent widths and line ratios of the optical emission lines, high-ionization UV lines, and the presence of a strong IR excess from heated dust.  We demonstrate that all of these properties are largely consistent with a hard ionization field produced by a compact, low metallicity starburst (see also examples from \citealp{fosbury:03, erb:10} and discussion by \citealp{hunt:10}).  Further evidence comes from the fact that the lens resolves two line-emitting components of \lensID, though extended narrow-line regions excited by nuclear activity have been observed \citep[e.g.,][]{unger:87}.  Both star-formation and AGN reach a peak in their activity at $z\sim2$ so robust identification and separation of the two contributions is critical for understanding their effect on galaxy evolution.

We conclude by noting that we will obtain a substantial sample of (unlensed) galaxies with similar, if not quite as extreme, properties to \lensID\ at $1.3 < z < 2.2$ (covering [\ion{O}{3}]+H$\beta$) in the full 3D-HST survey.  Without the factor of $\sim$25 lens magnification, such an object would have $m_\mathrm{F140W}=25.2$, where 3D-HST is sensitive to line equivalent widths $\gtrsim$1000~\AA\ \citep{brammer:3dhst}.  Indeed many such galaxies have recently been found with WFC3 \citep{atek:11, vanderwel:11}.  While individual unlensed objects will not allow such detailed study as performed here, statistical samples of these galaxies will offer insights into the metallicity and star-formation properties of the low-mass building blocks of galaxies observed today. 

%
%
\acknowledgements

We thank the referee, M. Malkan, for constructive comments that greatly improved the manuscript and R. Gavazzi and T. Treu for helpful discussions and for providing the LRIS spectrum of \lensID.  This publication makes use of data products from the Wide-field Infrared Survey Explorer, which is a joint project of the University of California, Los Angeles, and the Jet Propulsion Laboratory/California Institute of Technology, funded by the National Aeronautics and Space Administration.  Funding for this research was provided in part by the Marie Curie Actions of the European Commission (FP7-COFUND) and ERC grant HIGHZ no. 227749. 

{\it Facilities:} \facility{Hubble Space Telescope (WFC3)}

\bibliographystyle{apj}

\begin{thebibliography}{45}
\expandafter\ifx\csname natexlab\endcsname\relax\def\natexlab#1{#1}\fi

\bibitem[{{Abazajian} {et~al.}(2009){Abazajian}, {Adelman-McCarthy},
  {Ag{\"u}eros}, {Allam}, {Allende Prieto}, {An}, {Anderson}, {Anderson},
  {Annis}, {Bahcall}, {Bailer-Jones}, {Barentine}, {Bassett}, {Becker},
  {Beers}, {Bell}, {Belokurov}, {Berlind}, {Berman}, {Bernardi}, {Bickerton},
  {Bizyaev}, {Blakeslee}, {Blanton}, {Bochanski}, {Boroski}, {Brewington},
  {Brinchmann}, {Brinkmann}, {Brunner}, {Budav{\'a}ri}, {Carey}, {Carliles},
  {Carr}, {Castander}, {Cinabro}, {Connolly}, {Csabai}, {Cunha}, {Czarapata},
  {Davenport}, {de Haas}, {Dilday}, {Doi}, {Eisenstein}, {Evans}, {Evans},
  {Fan}, {Friedman}, {Frieman}, {Fukugita}, {G{\"a}nsicke}, {Gates},
  {Gillespie}, {Gilmore}, {Gonzalez}, {Gonzalez}, {Grebel}, {Gunn},
  {Gy{\"o}ry}, {Hall}, {Harding}, {Harris}, {Harvanek}, {Hawley}, {Hayes},
  {Heckman}, {Hendry}, {Hennessy}, {Hindsley}, {Hoblitt}, {Hogan}, {Hogg},
  {Holtzman}, {Hyde}, {Ichikawa}, {Ichikawa}, {Im}, {Ivezi{\'c}}, {Jester},
  {Jiang}, {Johnson}, {Jorgensen}, {Juri{\'c}}, {Kent}, {Kessler}, {Kleinman},
  {Knapp}, {Konishi}, {Kron}, {Krzesinski}, {Kuropatkin}, {Lampeitl},
  {Lebedeva}, {Lee}, {Lee}, {Leger}, {L{\'e}pine}, {Li}, {Lima}, {Lin}, {Long},
  {Loomis}, {Loveday}, {Lupton}, {Magnier}, {Malanushenko}, {Malanushenko},
  {Mandelbaum}, {Margon}, {Marriner}, {Mart{\'{\i}}nez-Delgado}, {Matsubara},
  {McGehee}, {McKay}, {Meiksin}, {Morrison}, {Mullally}, {Munn}, {Murphy},
  {Nash}, {Nebot}, {Neilsen}, {Newberg}, {Newman}, {Nichol}, {Nicinski},
  {Nieto-Santisteban}, {Nitta}, {Okamura}, {Oravetz}, {Ostriker}, {Owen},
  {Padmanabhan}, {Pan}, {Park}, {Pauls}, {Peoples}, {Percival}, {Pier}, {Pope},
  {Pourbaix}, {Price}, {Purger}, {Quinn}, {Raddick}, {Fiorentin}, {Richards},
  {Richmond}, {Riess}, {Rix}, {Rockosi}, {Sako}, {Schlegel}, {Schneider},
  {Scholz}, {Schreiber}, {Schwope}, {Seljak}, {Sesar}, {Sheldon}, {Shimasaku},
  {Sibley}, {Simmons}, {Sivarani}, {Smith}, {Smith}, {Smol{\v c}i{\'c}},
  {Snedden}, {Stebbins}, {Steinmetz}, {Stoughton}, {Strauss}, {Subba Rao},
  {Suto}, {Szalay}, {Szapudi}, {Szkody}, {Tanaka}, {Tegmark}, {Teodoro},
  {Thakar}, {Tremonti}, {Tucker}, {Uomoto}, {Vanden Berk}, {Vandenberg},
  {Vidrih}, {Vogeley}, {Voges}, {Vogt}, {Wadadekar}, {Watters}, {Weinberg},
  {West}, {White}, {Wilhite}, {Wonders}, {Yanny}, {Yocum}, {York}, {Zehavi},
  {Zibetti}, \& {Zucker}}]{sdss:dr7}
{Abazajian}, K.~N., {Adelman-McCarthy}, J.~K., {Ag{\"u}eros}, M.~A., {et~al.}
  2009, \apjs, 182, 543

\bibitem[{{Asplund} {et~al.}(2009){Asplund}, {Grevesse}, {Sauval}, \&
  {Scott}}]{asplund:09}
{Asplund}, M., {Grevesse}, N., {Sauval}, A.~J., {et~al.} 2009, \araa, 47, 481

\bibitem[{{Atek} {et~al.}(2011){Atek}, {Siana}, {Scarlata}, {Malkan},
  {McCarthy}, {Teplitz}, {Henry}, {Colbert}, {Bridge}, {Bunker}, {Dressler},
  {Fosbury}, {Hathi}, {Martin}, {Ross}, \& {Shim}}]{atek:11}
{Atek}, H., {Siana}, B., {Scarlata}, C., {et~al.} 2011, \apj, 743, 121

\bibitem[{{Brammer} {et~al.}(2012){Brammer}, {van Dokkum}, {Franx},
  {Fumagalli}, {Patel}, {Rix}, {Skelton}, {Kriek}, {Nelson}, {Schmidt},
  {Bezanson}, {da Cunha}, {Erb}, {Fan}, {F{\"o}rster Schreiber}, {Illingworth},
  {Labb{\'e}}, {Leja}, {Lundgren}, {Magee}, {Marchesini}, {McCarthy},
  {Momcheva}, {Muzzin}, {Quadri}, {Steidel}, {Tal}, {Wake}, {Whitaker}, \&
  {Williams}}]{brammer:3dhst}
{Brammer}, G.~B., {van Dokkum}, P.~G., {Franx}, M., {et~al.} 2012, \apjs, 200,
  13

\bibitem[{{Brinchmann} {et~al.}(2004){Brinchmann}, {Charlot}, {White},
  {Tremonti}, {Kauffmann}, {Heckman}, \& {Brinkmann}}]{brinchmann:04}
{Brinchmann}, J., {Charlot}, S., {White}, S.~D.~M., {et~al.} 2004, \mnras, 351,
  1151

\bibitem[{{Brinchmann} {et~al.}(2008){Brinchmann}, {Kunth}, \&
  {Durret}}]{brinchmann:08}
{Brinchmann}, J., {Kunth}, D., \& {Durret}, F. 2008, \aap, 485, 657

\bibitem[{{Bruzual} \& {Charlot}(2003)}]{bc:03}
{Bruzual}, G., \& {Charlot}, S. 2003, \mnras, 344, 1000

\bibitem[{{Calzetti} {et~al.}(2000){Calzetti}, {Armus}, {Bohlin}, {Kinney},
  {Koornneef}, \& {Storchi-Bergmann}}]{calzetti:00}
{Calzetti}, D., {Armus}, L., {Bohlin}, R.~C., {et~al.} 2000, \apj, 533, 682

\bibitem[{{Cooray} {et~al.}(2011){Cooray}, {Fu}, {Calanog}, {Wardlow}, {Chiu},
  {Kim}, {Smidt}, {Acquaviva}, {Ferguson}, {Faber}, {Galametz}, {Grogin},
  {Hartley}, {Kocevski}, {Koekemoer}, {Koo}, {Lucas}, {Moustakas}, \&
  {Newman}}]{cooray:11}
{Cooray}, A., {Fu}, H., {Calanog}, J., {et~al.} 2011, ArXiv e-prints

\bibitem[{{Dopita} \& {Sutherland}(2003)}]{dopita+sutherland}
{Dopita}, M.~A., \& {Sutherland}, R.~S. 2003, {Astrophysics of the diffuse
  universe}

\bibitem[{{Engelbracht} {et~al.}(2008){Engelbracht}, {Rieke}, {Gordon},
  {Smith}, {Werner}, {Moustakas}, {Willmer}, \& {Vanzi}}]{engelbracht:08}
{Engelbracht}, C.~W., {Rieke}, G.~H., {Gordon}, K.~D., {et~al.} 2008, \apj,
  678, 804

\bibitem[{{Erb} {et~al.}(2010){Erb}, {Pettini}, {Shapley}, {Steidel}, {Law}, \&
  {Reddy}}]{erb:10}
{Erb}, D.~K., {Pettini}, M., {Shapley}, A.~E., {et~al.} 2010, \apj, 719, 1168

\bibitem[{{Erb} {et~al.}(2006){Erb}, {Steidel}, {Shapley}, {Pettini}, {Reddy},
  \& {Adelberger}}]{erb:06}
{Erb}, D.~K., {Steidel}, C.~C., {Shapley}, A.~E., {et~al.} 2006, \apj, 646, 107

\bibitem[{{Finkelstein} {et~al.}(2011){Finkelstein}, {Papovich}, {Salmon},
  {Finlator}, {Dickinson}, {Ferguson}, {Giavalisco}, {Koekemoer}, {Reddy},
  {Bassett}, {Conselice}, {Dunlop}, {Faber}, {Grogin}, {Hathi}, {Kocevski},
  {Lai}, {Lee}, {McLure}, {Mobasher}, \& {Newman}}]{finkelstein:11}
{Finkelstein}, S.~L., {Papovich}, C., {Salmon}, B., {et~al.} 2011, ArXiv
  e-prints

\bibitem[{{Foreman-Mackey} {et~al.}(2012){Foreman-Mackey}, {Hogg}, {Lang}, \&
  {Goodman}}]{emcee}
{Foreman-Mackey}, D., {Hogg}, D.~W., {Lang}, D., {et~al.} 2012, ArXiv e-prints

\bibitem[{{Fosbury} {et~al.}(2003){Fosbury}, {Villar-Mart{\'{\i}}n},
  {Humphrey}, {Lombardi}, {Rosati}, {Stern}, {Hook}, {Holden}, {Stanford},
  {Squires}, {Rauch}, \& {Sargent}}]{fosbury:03}
{Fosbury}, R.~A.~E., {Villar-Mart{\'{\i}}n}, M., {Humphrey}, A., {et~al.} 2003,
  \apj, 596, 797

\bibitem[{{Griffith} {et~al.}(2011){Griffith}, {Tsai}, {Stern}, {Blain},
  {Eisenhardt}, {Harrison}, {Jarrett}, {Madsen}, {Stanford}, {Wright}, {Wu},
  {Wu}, \& {Yan}}]{griffith:11}
{Griffith}, R.~L., {Tsai}, C.-W., {Stern}, D., {et~al.} 2011, \apjl, 736, L22

\bibitem[{{Grogin} {et~al.}(2011){Grogin}, {Kocevski}, {Faber}, {Ferguson},
  {Koekemoer}, {Riess}, {Acquaviva}, {Alexander}, {Almaini}, {Ashby}, {Barden},
  {Bell}, {Bournaud}, {Brown}, {Caputi}, {Casertano}, {Cassata}, {Castellano},
  {Challis}, {Chary}, {Cheung}, {Cirasuolo}, {Conselice}, {Roshan Cooray},
  {Croton}, {Daddi}, {Dahlen}, {Dav{\'e}}, {de Mello}, {Dekel}, {Dickinson},
  {Dolch}, {Donley}, {Dunlop}, {Dutton}, {Elbaz}, {Fazio}, {Filippenko},
  {Finkelstein}, {Fontana}, {Gardner}, {Garnavich}, {Gawiser}, {Giavalisco},
  {Grazian}, {Guo}, {Hathi}, {H{\"a}ussler}, {Hopkins}, {Huang}, {Huang},
  {Jha}, {Kartaltepe}, {Kirshner}, {Koo}, {Lai}, {Lee}, {Li}, {Lotz}, {Lucas},
  {Madau}, {McCarthy}, {McGrath}, {McIntosh}, {McLure}, {Mobasher},
  {Moustakas}, {Mozena}, {Nandra}, {Newman}, {Niemi}, {Noeske}, {Papovich},
  {Pentericci}, {Pope}, {Primack}, {Rajan}, {Ravindranath}, {Reddy}, {Renzini},
  {Rix}, {Robaina}, {Rodney}, {Rosario}, {Rosati}, {Salimbeni}, {Scarlata},
  {Siana}, {Simard}, {Smidt}, {Somerville}, {Spinrad}, {Straughn}, {Strolger},
  {Telford}, {Teplitz}, {Trump}, {van der Wel}, {Villforth}, {Wechsler},
  {Weiner}, {Wiklind}, {Wild}, {Wilson}, {Wuyts}, {Yan}, \& {Yun}}]{grogin:11}
{Grogin}, N.~A., {Kocevski}, D.~D., {Faber}, S.~M., {et~al.} 2011, \apjs, 197,
  35

\bibitem[{{Hainline} {et~al.}(2011){Hainline}, {Shapley}, {Greene}, \&
  {Steidel}}]{hainline:11}
{Hainline}, K.~N., {Shapley}, A.~E., {Greene}, J.~E., {et~al.} 2011, \apj, 733,
  31

\bibitem[{{Hirashita} \& {Hunt}(2004)}]{hirashita:04}
{Hirashita}, H., \& {Hunt}, L.~K. 2004, \aap, 421, 555

\bibitem[{{Hunt} {et~al.}(2010){Hunt}, {Thuan}, {Izotov}, \&
  {Sauvage}}]{hunt:10}
{Hunt}, L.~K., {Thuan}, T.~X., {Izotov}, Y.~I., {et~al.} 2010, \apj, 712, 164

\bibitem[{{Izotov} {et~al.}(2011){Izotov}, {Guseva}, {Fricke}, \&
  {Henkel}}]{izotov:11}
{Izotov}, Y.~I., {Guseva}, N.~G., {Fricke}, K.~J., {et~al.} 2011, \aap, 536, L7

\bibitem[{{Izotov} {et~al.}(2009){Izotov}, {Guseva}, {Fricke}, \&
  {Papaderos}}]{izotov:09}
---. 2009, \aap, 503, 61

\bibitem[{{Izotov} {et~al.}(1997){Izotov}, {Lipovetsky}, {Chaffee}, {Foltz},
  {Guseva}, \& {Kniazev}}]{izotov:97}
{Izotov}, Y.~I., {Lipovetsky}, V.~A., {Chaffee}, F.~H., {et~al.} 1997, \apj,
  476, 698

\bibitem[{{Izotov} {et~al.}(2006){Izotov}, {Stasi{\'n}ska}, {Meynet}, {Guseva},
  \& {Thuan}}]{izotov:06}
{Izotov}, Y.~I., {Stasi{\'n}ska}, G., {Meynet}, G., {et~al.} 2006, \aap, 448,
  955

\bibitem[{{Kennicutt}(1998)}]{kennicutt:98}
{Kennicutt}, Jr., R.~C. 1998, \araa, 36, 189

\bibitem[{{Koekemoer} {et~al.}(2011){Koekemoer}, {Faber}, {Ferguson}, {Grogin},
  {Kocevski}, {Koo}, {Lai}, {Lotz}, {Lucas}, {McGrath}, {Ogaz}, {Rajan},
  {Riess}, {Rodney}, {Strolger}, {Casertano}, {Castellano}, {Dahlen},
  {Dickinson}, {Dolch}, {Fontana}, {Giavalisco}, {Grazian}, {Guo}, {Hathi},
  {Huang}, {van der Wel}, {Yan}, {Acquaviva}, {Alexander}, {Almaini}, {Ashby},
  {Barden}, {Bell}, {Bournaud}, {Brown}, {Caputi}, {Cassata}, {Challis},
  {Chary}, {Cheung}, {Cirasuolo}, {Conselice}, {Roshan Cooray}, {Croton},
  {Daddi}, {Dav{\'e}}, {de Mello}, {de Ravel}, {Dekel}, {Donley}, {Dunlop},
  {Dutton}, {Elbaz}, {Fazio}, {Filippenko}, {Finkelstein}, {Frazer}, {Gardner},
  {Garnavich}, {Gawiser}, {Gruetzbauch}, {Hartley}, {H{\"a}ussler},
  {Herrington}, {Hopkins}, {Huang}, {Jha}, {Johnson}, {Kartaltepe},
  {Khostovan}, {Kirshner}, {Lani}, {Lee}, {Li}, {Madau}, {McCarthy},
  {McIntosh}, {McLure}, {McPartland}, {Mobasher}, {Moreira}, {Mortlock},
  {Moustakas}, {Mozena}, {Nandra}, {Newman}, {Nielsen}, {Niemi}, {Noeske},
  {Papovich}, {Pentericci}, {Pope}, {Primack}, {Ravindranath}, {Reddy},
  {Renzini}, {Rix}, {Robaina}, {Rosario}, {Rosati}, {Salimbeni}, {Scarlata},
  {Siana}, {Simard}, {Smidt}, {Snyder}, {Somerville}, {Spinrad}, {Straughn},
  {Telford}, {Teplitz}, {Trump}, {Vargas}, {Villforth}, {Wagner}, {Wandro},
  {Wechsler}, {Weiner}, {Wiklind}, {Wild}, {Wilson}, {Wuyts}, \&
  {Yun}}]{koekemoer:11}
{Koekemoer}, A.~M., {Faber}, S.~M., {Ferguson}, H.~C., {et~al.} 2011, \apjs,
  197, 36

\bibitem[{{Labb{\'e}} {et~al.}(2006){Labb{\'e}}, {Bouwens}, {Illingworth}, \&
  {Franx}}]{labbe:06}
{Labb{\'e}}, I., {Bouwens}, R., {Illingworth}, G.~D., {et~al.} 2006, \apjl,
  649, L67

\bibitem[{{Malkan} {et~al.}(1996){Malkan}, {Teplitz}, \& {McLean}}]{malkan:96}
{Malkan}, M.~A., {Teplitz}, H., \& {McLean}, I.~S. 1996, \apjl, 468, L9

\bibitem[{{Mannucci} {et~al.}(2009){Mannucci}, {Cresci}, {Maiolino}, {Marconi},
  {Pastorini}, {Pozzetti}, {Gnerucci}, {Risaliti}, {Schneider}, {Lehnert}, \&
  {Salvati}}]{mannucci:09}
{Mannucci}, F., {Cresci}, G., {Maiolino}, R., {et~al.} 2009, \mnras, 398, 1915

\bibitem[{{Morales-Luis} {et~al.}(2011){Morales-Luis}, {S{\'a}nchez Almeida},
  {Aguerri}, \& {Mu{\~n}oz-Tu{\~n}{\'o}n}}]{morales-luis:11}
{Morales-Luis}, A.~B., {S{\'a}nchez Almeida}, J., {Aguerri}, J.~A.~L., {et~al.}
  2011, \apj, 743, 77

\bibitem[{{Overzier} {et~al.}(2009){Overzier}, {Heckman}, {Tremonti}, {Armus},
  {Basu-Zych}, {Gon{\c c}alves}, {Rich}, {Martin}, {Ptak}, {Schiminovich},
  {Ford}, {Madore}, \& {Seibert}}]{overzier:09}
{Overzier}, R.~A., {Heckman}, T.~M., {Tremonti}, C., {et~al.} 2009, \apj, 706,
  203

\bibitem[{{Peng} {et~al.}(2002){Peng}, {Ho}, {Impey}, \& {Rix}}]{peng:02}
{Peng}, C.~Y., {Ho}, L.~C., {Impey}, C.~D., {et~al.} 2002, \aj, 124, 266

\bibitem[{{Reines} {et~al.}(2008){Reines}, {Johnson}, \& {Hunt}}]{reines:08}
{Reines}, A.~E., {Johnson}, K.~E., \& {Hunt}, L.~K. 2008, \aj, 136, 1415

\bibitem[{{Thuan} {et~al.}(1997){Thuan}, {Izotov}, \& {Lipovetsky}}]{thuan:97}
{Thuan}, T.~X., {Izotov}, Y.~I., \& {Lipovetsky}, V.~A. 1997, \apj, 477, 661

\bibitem[{{Thuan} {et~al.}(1999){Thuan}, {Sauvage}, \& {Madden}}]{thuan:99}
{Thuan}, T.~X., {Sauvage}, M., \& {Madden}, S. 1999, \apj, 516, 783

\bibitem[{{Todini} \& {Ferrara}(2001)}]{todini:01}
{Todini}, P., \& {Ferrara}, A. 2001, \mnras, 325, 726

\bibitem[{{Tremonti} {et~al.}(2004){Tremonti}, {Heckman}, {Kauffmann},
  {Brinchmann}, {Charlot}, {White}, {Seibert}, {Peng}, {Schlegel}, {Uomoto},
  {Fukugita}, \& {Brinkmann}}]{tremonti:04}
{Tremonti}, C.~A., {Heckman}, T.~M., {Kauffmann}, G., {et~al.} 2004, \apj, 613,
  898

\bibitem[{{Trump} {et~al.}(2011){Trump}, {Weiner}, {Scarlata}, {Kocevski},
  {Bell}, {McGrath}, {Koo}, {Faber}, {Laird}, {Mozena}, {Rangel}, {Yan},
  {Yesuf}, {Atek}, {Dickinson}, {Donley}, {Dunlop}, {Ferguson}, {Finkelstein},
  {Grogin}, {Hathi}, {Juneau}, {Kartaltepe}, {Koekemoer}, {Nandra}, {Newman},
  {Rodney}, {Straughn}, \& {Teplitz}}]{trump:11}
{Trump}, J.~R., {Weiner}, B.~J., {Scarlata}, C., {et~al.} 2011, \apj, 743, 144

\bibitem[{{Tu} {et~al.}(2009){Tu}, {Gavazzi}, {Limousin}, {Cabanac},
  {Marshall}, {Fort}, {Treu}, {P{\'e}llo}, {Jullo}, {Kneib}, \&
  {Sygnet}}]{tu:09}
{Tu}, H., {Gavazzi}, R., {Limousin}, M., {et~al.} 2009, \aap, 501, 475

\bibitem[{{Unger} {et~al.}(1987){Unger}, {Pedlar}, {Axon}, {Whittle}, {Meurs},
  \& {Ward}}]{unger:87}
{Unger}, S.~W., {Pedlar}, A., {Axon}, D.~J., {et~al.} 1987, \mnras, 228, 671

\bibitem[{{van der Wel} {et~al.}(2011){van der Wel}, {Straughn}, {Rix},
  {Finkelstein}, {Koekemoer}, {Weiner}, {Wuyts}, {Bell}, {Faber}, {Trump},
  {Koo}, {Ferguson}, {Scarlata}, {Hathi}, {Dunlop}, {Newman}, {Dickinson},
  {Jahnke}, {Salmon}, {de Mello}, {Kocevski}, {Lai}, {Grogin}, {Rodney}, {Guo},
  {McGrath}, {Lee}, {Barro}, {Huang}, {Riess}, {Ashby}, \&
  {Willner}}]{vanderwel:11}
{van der Wel}, A., {Straughn}, A.~N., {Rix}, H.-W., {et~al.} 2011, \apj, 742,
  111

\bibitem[{{Wright} {et~al.}(2010){Wright}, {Eisenhardt}, {Mainzer}, {Ressler},
  {Cutri}, {Jarrett}, {Kirkpatrick}, {Padgett}, {McMillan}, {Skrutskie},
  {Stanford}, {Cohen}, {Walker}, {Mather}, {Leisawitz}, {Gautier}, {McLean},
  {Benford}, {Lonsdale}, {Blain}, {Mendez}, {Irace}, {Duval}, {Liu}, {Royer},
  {Heinrichsen}, {Howard}, {Shannon}, {Kendall}, {Walsh}, {Larsen}, {Cardon},
  {Schick}, {Schwalm}, {Abid}, {Fabinsky}, {Naes}, \& {Tsai}}]{wright:wise}
{Wright}, E.~L., {Eisenhardt}, P.~R.~M., {Mainzer}, A.~K., {et~al.} 2010, \aj,
  140, 1868

\bibitem[{{Wuyts} {et~al.}(2012){Wuyts}, {Rigby}, {Sharon}, \&
  {Gladders}}]{ewuyts:12}
{Wuyts}, E., {Rigby}, J.~R., {Sharon}, K., {et~al.} 2012, ArXiv e-prints

\bibitem[{{Yuan} \& {Kewley}(2009)}]{yuan:09}
{Yuan}, T.-T., \& {Kewley}, L.~J. 2009, \apjl, 699, L161

\end{thebibliography}

\end{document}